# About existence of stationary points for the Arnold-Beltrami-Childress (*ABC*) flow

**Sergey V. Ershkov**

Institute for Time Nature Explorations,

M.V. Lomonosov's Moscow State University,

Leninskie gory, 1-12, Moscow 119991, Russia

e-mail: sergej-ershkov@yandex.ru

The existence of stationary points for the dynamical system of *ABC*-flow is considered. The *ABC*-flow, a three-parameter velocity field that provides a simple stationary solution of Euler's equations in three dimensions for incompressible, inviscid fluid flows, is the prototype for the study of *turbulence* (it provides a simple example of dynamical chaos).

But, nevertheless, between the chaotic trajectories of the appropriate solutions of such a system we can reveal the stationary points, the deterministic basis among the chaotic behaviour of ABC-flow dynamical system. It has been proved the existence of 1 point for two partial cases of parameters $\{A, B, C\}$: 1) $A = B = 1$; 2) $C = 1$ ($A^2 + B^2 = 1$). Moreover, dynamical system of *ABC*-flow allows 3 points of such a type, depending on the meanings of parameters $\{A, B, C\}$.

# 1. Introduction, the *ABC*-flow.

The Arnold-Beltrami-Childres flow [1-2], a well known helical steady solution of Euler equations for *ideal* incompressible flow of Newtonian fluids, should be presented in the Cartesian coordinates as below (*A*, *B*, *C* = const):

$$u_1 = A\sin z + C\cos y, \quad u_2 = B\sin x + A\cos z, \quad u_3 = C\sin y + B\cos x \qquad (1.1)$$

- where $\boldsymbol{u} = \{u_1, u_2, u_3\}$ is the vector of flow velocity field, a particular simple (*steady*) case of *helical* flow:

$$\vec{w} = \alpha \cdot \vec{u} \implies \vec{u} \times \vec{w} = \vec{0},$$

- here $\alpha$ is the constant coefficient, given by the initial conditions ($\alpha \neq 0$). Such a type of solution (1.1) is also known as Beltrami flow [3-4], i.e. a fluid motion in which the *vorticity* vector is parallel to the velocity vector at every point of the fluid.

As for the domain in which the flow occurs, let us consider the periodic boundary conditions and the limited domain.

If we remember the originating of denotation for the components of velocity field:

$$u_1 = \frac{dx}{dt}, \quad u_2 = \frac{dy}{dt}, \quad u_3 = \frac{dz}{dt},$$

- it should yield a system of ordinary differential equations as below

$$\begin{cases} \dfrac{dx}{dt} = A\sin z + C\cos y, \\[6pt] \dfrac{dy}{dt} = B\sin x + A\cos z, \\[6pt] \dfrac{dz}{dt} = C\sin y + B\cos x, \end{cases} \qquad (1.2)$$

- which is proved to have not an analytical solutions, but moreover it reveals a dynamical chaos among the trajectories of appropriate solutions of such a system [5-6].

The *ABC*-flow, a three-parameter velocity field that provides a simple stationary solution of Euler's equations in three dimensions for incompressible, inviscid fluid flows, can be considered to be a prototype for the study of turbulence: the ABC-flow provides a simple example of dynamical chaos, in spite of the simple analytical expression for each of the components of a solution [7].

## 2. Analysis of the existence of stationary points in ABC-flow.

Let us explore the fixed (stationary) points of *ABC*-flow system (1.2), which could be associated with the local *stability* of the dynamical trajectories for such a system. It means that there should be valid appropriate conditions $dx/dt = dy/dt = dz/dt = 0$ [8]; so, we obtain from the equations of *ABC*-flow dynamical system (1.2) as below:

$$\begin{cases} C\cos y = -A\sin z, \\ B\sin x = -A\cos z, \\ B\cos x = -C\sin y. \end{cases} \quad (2.1)$$

System of Eqs. (2.1) yields *three* types of solutions, depending on the meanings of parameters $\{A, B, C\}$, as presented below

$$C^2(\cos^2 y - \sin^2 y) + B^2 = A^2 \Rightarrow y = \frac{1}{2}\arccos\left(\frac{A^2 - B^2}{C^2}\right), \quad -C^2 \leq A^2 - B^2 \leq C^2 \quad (2.2)$$

$$A^2(\cos^2 z - \sin^2 z) + C^2 = B^2 \Rightarrow z = \frac{1}{2}\arccos\left(\frac{B^2 - C^2}{A^2}\right), \quad -A^2 \leq B^2 - C^2 \leq A^2 \quad (2.3)$$

$$B^2(\cos^2 x - \sin^2 x) + A^2 = C^2 \Rightarrow x = \frac{1}{2}\arccos\left(\frac{C^2 - A^2}{B^2}\right), \quad -B^2 \leq C^2 - A^2 \leq B^2 \quad (2.4)$$

For the solutions of a type (2.2) we could obtain from system (2.1) as below

$$x = \pm \arccos\left(\mp \frac{C}{B}\sin\left(\frac{1}{2}\arccos\left(\frac{A^2 - B^2}{C^2}\right)\right)\right), \quad z = -\arcsin\left(\frac{C}{A}\cos\left(\frac{1}{2}\arccos\left(\frac{A^2 - B^2}{C^2}\right)\right)\right) \quad (2.5)$$

If we assume $A = B = 1$, the last expressions for $x$, $z$ could be simplified properly

$$x = \pm \arccos\left(\mp \frac{C}{\sqrt{2}}\right), \quad z = -\arcsin\left(\frac{C}{\sqrt{2}}\right), \quad -\sqrt{2} \leq C \leq \sqrt{2} \quad (2.6)$$

Solutions (2.2), (2.6) should be substituted (under assumptions $A = B = 1$) to the system (2.1) for the resulting checking of such a solution.

3. **Analysis for the existence of stationary points in ABC-flow.**

So, if there exists a fixed (stationary) point of *ABC*-flow dynamical system (1.2) for the case $A = B = 1$, the condition below should be valid for the range of meanings of parameter *C*:

$$\pm \sin\left(\arccos\left(\mp \frac{C}{\sqrt{2}}\right)\right) = -\cos\left(\arcsin\left(\frac{C}{\sqrt{2}}\right)\right) \quad (3.1)$$

- which is obviously valid for all the range of meanings of parameter *C* from inequality, shown in (2.6); besides, we should choose the sign "-" in a left part of expression (3.1) above. Indeed, let us present the appropriate plots of the left and right parts of Eq. (3.1) at Fig.1-2 below:

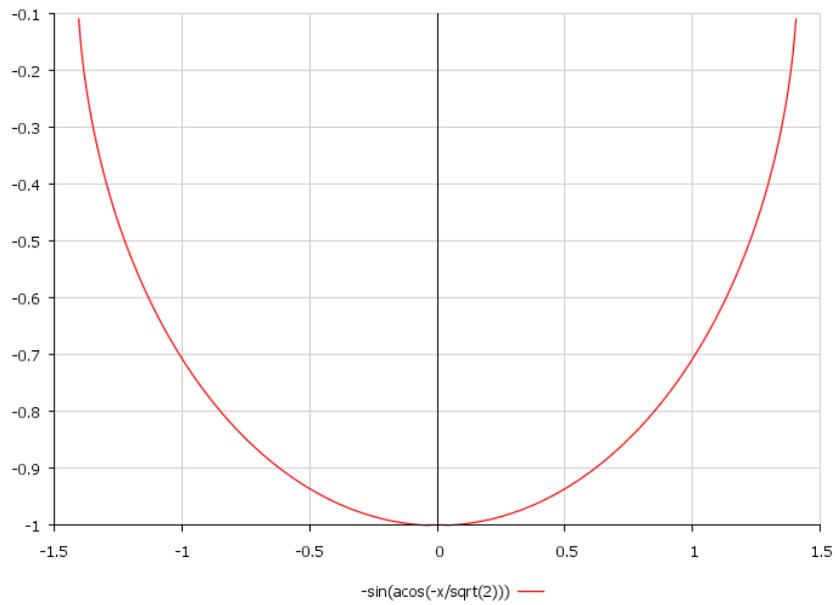

Fig.1. Plot of the left part of Eq. (3.1), here we denote $C$ = x.

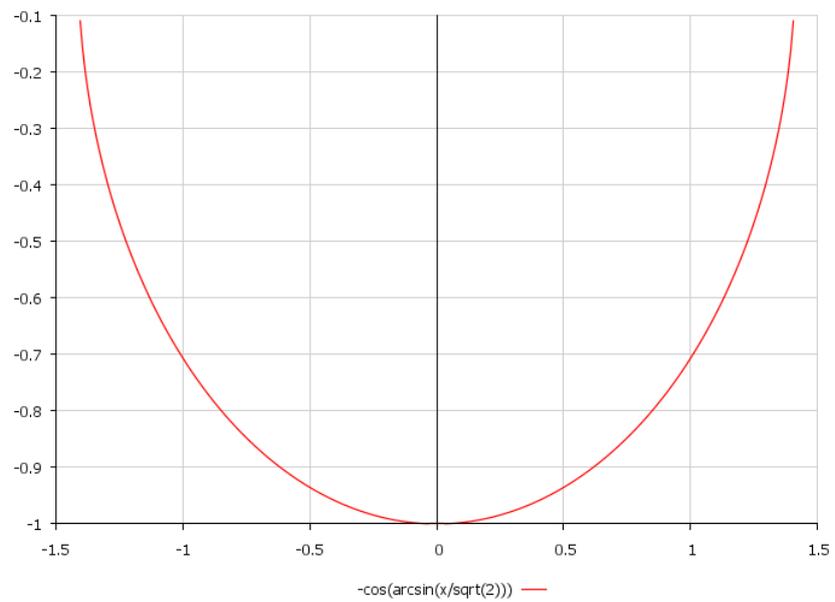

Fig.2. Plot of the right part of Eq. (3.1), here we denote $C$ = x.

In general case (2.5), if there exist a fixed (stationary) points of *ABC*-flow dynamical system (1.2), the condition below should be valid for the appropriate ranges of meanings of parameters $\{A, B, C\}$:

$$\pm B \sin\left(\arccos\left(\mp \frac{C}{B} \sin\left(\frac{1}{2} \arccos\left(\frac{A^2 - B^2}{C^2}\right)\right)\right)\right) =$$

$$= -A \cos\left(\arcsin\left(\frac{C}{A} \cos\left(\frac{1}{2} \arccos\left(\frac{A^2 - B^2}{C^2}\right)\right)\right)\right), \qquad (3.2)$$

- where we could choose $C = 1$ (moreover, we could choose any meanings of $C$ which should be *a scale-parameter* for the meanings of parameters $\{A, B\}$ in (2.1)).

Let us present the appropriate plots of the left and right parts of Eq. (3.2), at Fig.3-4 below (we should choose the sign "-" in a left part of expression (3.2) above):

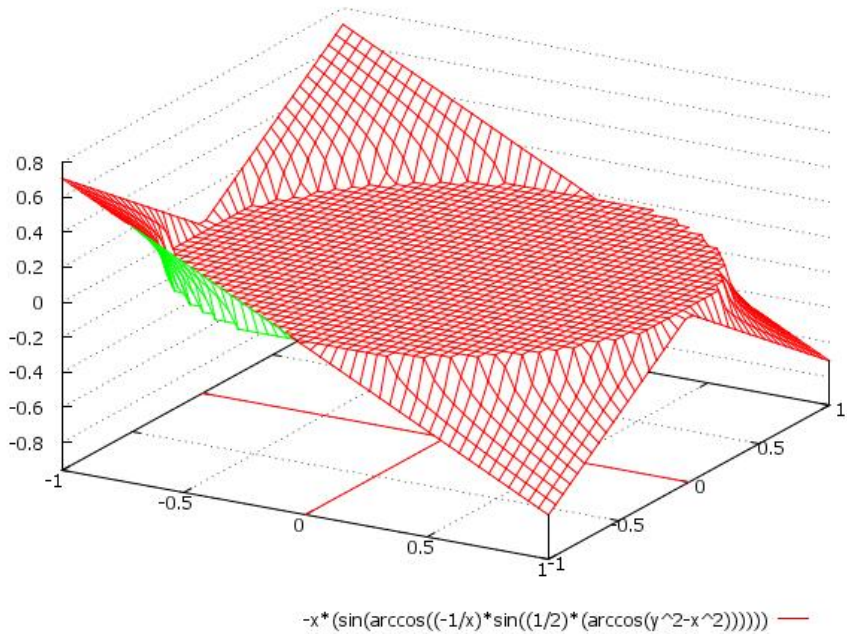

Fig.3. Plot of the left part of Eq. (3.2), here we denote $B = x$, $A = y$.

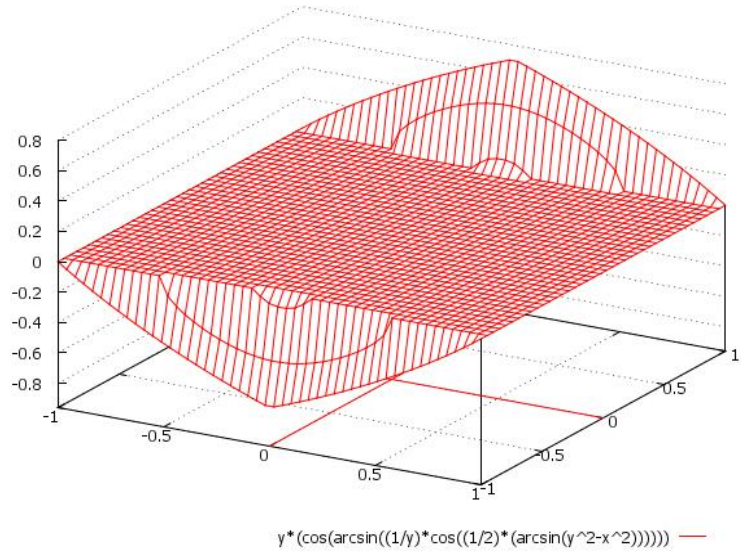

Fig.4. Plot of the right part of Eq. (3.2), here we denote $B = $ x, $A = $ y.

We can see at the Figs. 3-4 the only surface for the coinciding of the left and right parts of Eq. (3.2), which should apparently be the intersecting circle (see Fig.3) at zero plane:

$$-B\sin\left(\arccos\left(-\frac{1}{B}\sin\left(\frac{1}{2}\arccos(A^2 - B^2)\right)\right)\right) = 0 =$$

$$= -A\cos\left(\arcsin\left(\frac{1}{A}\cos\left(\frac{1}{2}\arccos(A^2 - B^2)\right)\right)\right), \qquad (3.3)$$

Moreover, we could obviously assume that an equality below is valid for such an intersecting circle

$$A^2 + B^2 = 1, \qquad (3.4)$$

- for which, the inequalities from (2.2) should be obviously valid as shown below:

$$-1 \leq A^2 - B^2 \leq 1 \quad \Rightarrow \quad (B^2 - 1) \leq A^2 \leq (B^2 + 1), \quad \Rightarrow \quad 0 \leq B^2 \leq 1.$$

So, using Eq. (3.4), we obtain from (3.3)

$$\begin{cases} \sin\left(\dfrac{1}{2}\arccos(A^2 - B^2)\right) = -B \\ \cos\left(\dfrac{1}{2}\arccos(A^2 - B^2)\right) = \pm A \end{cases} \quad (3.5)$$

- where (3.5) is obviously valid for chosen meanings of $\{A, B\}$ according to (3.4).

### 4. **Discussion.**

The main result which should be outlined is the existence of stationary points for the dynamical system (1.2) of *ABC*-flow.

The *ABC*-flow, a three-parameter velocity field that provides a simple stationary solution of Euler's equations in three dimensions for incompressible, inviscid fluid flows, can be considered to be a prototype for the study of *turbulence* - the ABC-flow provides a simple example of dynamical chaos [7].

But, nevertheless, between the chaotic trajectories of the appropriate solutions of such a system we can reveal the stationary points, the stationary kernel or deterministic basis among the chaotic behaviour of ABC-flow dynamical system.

If we assume $A = B = 1$, the expressions for coordinates $x$, $y$, $z$ (just for one of 3 types of solution (2.2)-(2.4)) should be as below

$$x = -\arccos\left(-\dfrac{C}{\sqrt{2}}\right), \quad y = -\dfrac{\pi}{4}, \quad z = -\arcsin\left(\dfrac{C}{\sqrt{2}}\right), \quad -\sqrt{2} \leq C \leq \sqrt{2} \quad (4.1)$$

- so, at least 1 point is possible in case $A = B = 1$ for each $C$ from the range of meanings in (4.1), just for one of 3 types of solution (2.2)-(2.4).

If we assume $C = 1$, the expressions for coordinates $x$, $y$, $z$ should be as below

$$x = -\arccos\left(-\frac{1}{B}\sin\left(\frac{1}{2}\arccos(A^2 - B^2)\right)\right),$$

$$y = -\frac{1}{2}\arccos(A^2 - B^2) \qquad (4.2)$$

$$z = -\arcsin\left(\frac{1}{A}\cos\left(\frac{1}{2}\arccos(A^2 - B^2)\right)\right)$$

- where according to (3.4) we should choose

$$A^2 + B^2 = 1,$$

- so, at least 1 point is also possible in case $C = 1$ for each meaning of $\{A, B\}$ above, just for one of 3 types of solution (2.2)-(2.4).

## 5. Conclusion.

The existence of stationary points for the dynamical system of *ABC*-flow is considered. The *ABC*-flow, a three-parameter velocity field that provides a simple stationary solution of Euler's equations in three dimensions for incompressible, inviscid fluid flows, is the prototype for the study of *turbulence* (it provides a simple example of dynamical chaos).
But, nevertheless, between the chaotic trajectories of the appropriate solutions of such a system we can reveal the stationary points, the deterministic basis among the chaotic behaviour of ABC-flow dynamical system. It has been proved the existence of 1 point for two partial cases of parameters $\{A, B, C\}$: 1) $A = B = 1$; 2) $C = 1$ ($A^2 + B^2 = 1$). Moreover, dynamical system of *ABC*-flow allows 3 points of such a type, depending on the meanings of parameters $\{A, B, C\}$.